\documentclass[preprint,showpacs,preprintnumbers,amsmath,amssymb,prb]{revtex4}
\usepackage{graphicx}
\begin{document}

\title{Transverse acoustic nature of the excess of vibrational
       states in vitreous silica}

\author{
        O.~Pilla$^1$, S.~Caponi$^2$, A.~Fontana$^1$,
        M.~Montagna$^1$, F.~Rossi$^1$,\\ G.~Viliani$^1$,
        L.~Angelani$^3$, G.~Ruocco$^3$, G.~Monaco$^4$, F.~Sette$^4$
       }

\affiliation{
         $^1$INFM and Dipartimento di Fisica,
         Universit\`a di Trento, 38050 Povo, Trento, Italy \\
         $^2$INFM and Dipartimento di Fisica, Universit\`a
         dell'Aquila, I-67100, L'Aquila, Italy \\
         $^3$INFM and Dipartimento di Fisica, Universit\`a
          di Roma {\it La Sapienza}, 00185 Roma, Italy \\
         $^4$European Synchrotron Radiation Facility,
         BP 220, 38043 Grenoble,
         France
         }

\date{\today}

\begin{abstract}
We present a numerical simulation study of the density-dependence
($\rho$=2.2$\div$4.0 g/cm$^3$) of the high-energy collective
dynamics in vitreous silica at mesoscopic wavevectors
($Q$=1$\div$18 nm$^{-1}$). The density-dependence of the
longitudinal and transverse current spectra provides evidence that
the excess modes observed in the density of states of this and
many other glasses, i.e. the Boson Peak, arises from the high-$Q$
limit of the quasi-{\it transverse} acoustic branch. This
conclusion emerges from the comparison of the numerical results
with the experimentally observed energy-shift and intensity
variation of the Boson Peak  with increasing density.
\end{abstract}


\maketitle

Insulating disordered solids, when compared to their crystalline
counterparts, exhibit some common peculiarities in their
low-temperature thermal properties and low-energy spectroscopic
features \cite{phill,phimaggen}, and in particular (i) a larger
specific heat at temperatures up to $\approx$1 K, ascribed to
tunneling processes \cite{anderson,phillips}, (ii) a much smaller
thermal conductivity, which also shows a plateau in the
temperature range $\approx$1$\div$10 K \cite{phill}, (iii) a
quasi-elastic light- and neutron-scattering intensity, and, most
important, (iv) an excess of modes in the vibrational density of
states, known as the Boson Peak (BP). An unambiguous
understanding of the origin of these extra modes, and of their
possible relation with other reported anomalies, is still
missing, in spite of the extensive research effort primed by the
pioneering work of Buchenau et al. \cite{buc86} and continued by
many authors
\cite{ben96,for96,dellanna,duval,schirm,tarask97,fontana99,pilla,tarask01}.
No definite conclusion can yet be drawn as to the nature of the
vibrational eigenvectors of the modes responsible for the BP:
according to different authors, in fact, they are either
spatially localized \cite{buc86,for96}, spatially delocalized and
propagating \cite{ben96,dellanna,pilla}, spatially delocalized
but diffusive in character \cite{feldman}. Even more important,
still we do not completely understand why should disorder
accumulate vibrational eigenvalues in the same energy region, in
such a broad variety of chemically and physically different
materials. Some recent theoretical works have been devoted to this
subject, and among them we recall (i) the work of Elliott and
coworkers, who assigned the BP in glasses to the lowest-energy
van Hove singularity of the corresponding crystal
\cite{tarask01,elliotpm02}; (ii) the work of Grigera et al., who
interpreted the BP as the precursor of the dynamical instability
expected in a disordered structure as function of density
\cite{Grigera}; and (iii) the work of G\"otze and Mayr and that
of Schilling et al., who obtained a spectral feature recalling
the BP within a mode-coupling-like description of the high
frequency dynamics of a model glass \cite{Goetze,Schilling}.

In this Letter, we present a Molecular Dynamics (MD) study of the
high-frequency dynamics in vitreous silica (v-SiO$_2$) at
different densities in the range $\rho$=2.2$\div$4.0 g/cm$^{3}$.
We demonstrate that the BP in vitreous silica arise from the high
$Q$ portion of the transverse acoustic branch. Our results
quantitatively support and better specify i) the Elliott and
coworkers' proposal \cite{tarask01,elliotpm02} that the BP is
connected to the van Hove singularity of crystalline quartz; and
ii) the Buchenau finding \cite{buc86} that the modes at the BP are
local rotation of SiO$_2$ tetrahedra.

Experimentally, in SiO$_2$ at room pressure the BP appears as a
broad peak in the plot of the Debye-normalized density of states
$g(E)/E^2$, centered at $\approx$5 meV and having a width of
$\approx$ 5 meV (Fig.1a). By increasing the sample density up to
$\rho$=4.0 g/cm$^{3}$, what is observed
\cite{sugai66,inamura,inabuche,jund} is a high-energy shift of the
peak and its concomitant intensity decrease (Fig. 1b,c). In
Fig.~1, the lines are experimental data, while symbols are the
results of the present simulations (see below), and we would like
to stress the excellent agreement between the two. It is important
to realize that the excess of modes in the density of state itself
$g(E)$, corresponding to the peak at $\approx$5 meV in $g(E)/E^2$
of Fig.~1, is actually centered (at room pressure) around
$\approx$ 15 meV, the shift being due to the $1/E^2$ factor and to
the very slow decay of the high frequency tail of the peak in
Fig.~1 .
With respect to this point, it is important to underline that -in
the high $Q$ limit and in the one-excitation (one-phonon)
approximation- the dynamic structure factor $S(Q,E)$ becomes
proportional to $g(E)/E^2$, and hence the longitudinal current
spectrum $C_L(Q,E)$=$E^2/Q^2S(Q,E)$ becomes $\propto$ $g(E)$.
Therefore, the modes giving rise to the BP must be searched, in
the current spectra, around 15 meV. This observation matches with
the high $Q$ panels of Fig.~2, where the computed currents  (vide
infra) show a broad bump at about this energy value.

The investigated system consists of 680 SiO$_{2}$ units ($N$=2040
ions), enclosed in cubic boxes of different lengths (from
$L$=3.1359 nm, corresponding to $\rho=2.2\,\,g/cm^{3}$ for the
glass at room pressure, down to $L$=2.5693 nm corresponding to a
density of 4.0 $g/cm^{3}$), with periodic boundary conditions. The
ions interact through the BKS \cite{vanbe} two-body interaction
potential; the long-range interaction was treated by the Ewald sum
technique. As it has already been demonstrated, this system
reproduce quantitatively the high frequency dynamics of vitreous
silica \cite{dellanna}. The glass configuration at room pressure
was obtained by standard MD methods for lowering the temperature
down to $300$ K starting from a well equilibrated liquid
configuration at $T\!=\!6000$ K, followed by a conjugate gradient
geometrical minimization on the potential-energy hypersurface for
an accurate location of the minimum. Such minimum configuration
was taken as the starting point to generate a series of compressed
systems. At each compression step, the box size was scaled by
$\approx$1.5\%, then the system was made to relax, and the new
minimum configuration was searched by the conjugate gradient
method. This procedure was repeated until the final density of 4.0
$g/cm^{3}$, corresponding to a sample under an hydrostatic
pressure of about 35 GPa \cite{zha}, was reached. A complete study
of the structural and dynamical changes occurring during the
compression will be reported in a forthcoming paper. We focus here
on the changes in the high frequency dynamics that take place as
the density is increased.

The vibrational dynamics in the minimum configurations was
computed in the harmonic approximation by diagonalizing the
dynamical matrix, to obtain the eigenvalues ($E_p$) and
eigenvectors ($e_p(i)$) of the $p$-th normal mode ($p=1 \div 3N$).
From these quantities all vibrational characteristics can be
derived. In particular, we have computed the density of states
$g(E)$ and the longitudinal and transverse currents spectra
($C(Q,E)$) which, in the one-excitation approximation, are given
by:
\begin{equation}
\label{corr} C^{\eta}_{_{\alpha\beta}}(Q,E) = [K_BT/\sqrt{M_\alpha
M_\beta}] \; \Sigma_p W^\eta_p(Q) \delta(E-E_p) \nonumber
\end{equation}
where $\eta=L,T$, $\alpha,\beta$ indicate Si and O, and
$W^{\eta}_p(Q)$  is the spatial power spectrum of the
(longitudinal or transverse) component of the eigenvectors:
\begin{eqnarray}
\label{eqw}
W^L_p(Q)&=&|\Sigma_i (\hat Q \cdot \bar e_p(i)) \exp{{(i \bar Q
\cdot \bar R_i )} |^2} \\
\nonumber
W^T_p(Q)&=&|\Sigma_i (\hat Q \times \bar e_p(i)) \exp{{(i \bar Q
\cdot \bar R_i )} |^2}.
\nonumber
\end{eqnarray}
Here $\hat Q$=$\bar Q/|Q|$.

The calculated $g(E)/E^2$ are reported in Fig.~1 together with the
corresponding experimental curves. The two sets of data compare
favorably to each other, indicating the suitability of the
employed potential model to follow the density dependence of the
high frequency dynamics. In Fig.~2 we report longitudinal and
transverse current spectra at selected $Q$ values in the
uncompressed sample ($\rho $=2.2 $g/cm^{3}$). For $Q$ values
larger than about 8 nm$^{-1}$, both $C^{^{L}}(Q,E)$ and
$C^{^{T}}(Q,E)$ show two distinct excitation maxima, a feature
that becomes more and more evident at increasing $Q$. The
excitation at higher energy disperses with $Q$ and is observed at
all $Q$ values in the longitudinal current spectra, while it shows
up as a weak shoulder in the transverse current spectra only at
$Q>10$ nm$^{-1}$. In agreement with previous findings
\cite{dellanna,horbach,taraskin98}, we assign this feature to the
longitudinal sound-like branch. The behaviour of the low-energy
excitation is in some sense complementary: it is always present in
the transverse current spectra, while it appears in the
longitudinal current spectra only at $Q>8$~nm$^{-1}$. At small
$Q$, the low-energy peak disperses with a sound velocity of
$\approx 3800$ $m/s$ (appropriate for the transverse sound mode),
and becomes almost non-dispersing at $Q>8$ nm$^{-1}$ (Fig. 3). We
will call this low-energy feature -which is the main feature in
the transverse current spectra- the transverse acoustic mode.

The presence of the signature of transverse dynamics in the {\it
longitudinal} current spectra, and vice-versa, is only apparently
surprising. Indeed, the polarization character of the modes (which
is better and better defined at increasing wavelength, i.~e. when
the vibration sees the medium as an elastic continuum) becomes ill
defined at short wavelengths ({\it mixing} phenomenon). This is at
the basis of the growth of peaks associated to the opposite
polarization modes in the current spectra, and of the increased
visibility of these peaks at increasing $Q$ values.
In Fig. 3 we report the computed dispersion curves of transverse
excitations (open symbols) for samples of three densities
($\rho=2.2, \,2.7,\, 4.0$ g/cm$^{3}$), and of longitudinal
excitations (crossed circles) of the uncompressed system \cite{nota2}.
In the same figure (full symbols), are also reported -at the available £Q£
values, the experimental excitation energies (maxima of the
current spectra) of the uncompressed SiO$_2$ glass, as measured by
IXS and INS \cite{fontana,righetti}. The agreement between MD and
experimental data clearly indicates that the peaks observed in the
INS experiment at $Q$ larger than 8 nm$^{-1}$ \cite{for96}, must
be associated with the transverse dynamics, which appears in the
measured (longitudinal) spectra due to the mixing.

In Fig.~3, we also observe a strong positive dispersion of the
velocity of the (longitudinal) sound waves, which continue to
propagate at energy well above the BP energy. This dispersion,
observable also in MD simulations of Lennard-Jones glasses
\cite{relax-harm} and in previous v-SiO$_2$ calculations
\cite{horbach}, can be due to the interaction of the sound waves
with a relaxation process \cite{relax-harm} or with other modes.
The investigation of the origin of the positive dispersion of
$v_L$, and is possible interplay with the BP modes, is beyond the
purpose of the present work.

For our present purposes, the first important result emerging from
Fig.~3 is that the "transverse" branches (open symbols) in low-
and intermediate-density samples, at large $Q$'s, flatten at high
$Q$'s to an energy value which increases with increasing density
($\approx$15 meV at room condition and $Q$=15 nm$^{-1}$). The
density of vibrational states ($g(E)$), associated with the
branches which flatten, will have an excess of modes with respect
to the Debye behaviour at these energies, reminiscent of the van
Hove singularity of the corresponding crystal \cite{tarask01}. As
mentioned, the corresponding peak in $g(E)/E^2$, will be
red-shifted at lower energies ($\approx$5 meV at room condition).
Therefore, we conclude that the Boson peak originates from the
modes associated to the flat portion of the acoustic transverse
dispersion curve.

This result is in agreement with the recent theoretical work of
S.~Taraskin et al., who associate the BP to the glassy counterpart
of the lowest-energy van Hove singularity of the corresponding
crystalline structure \cite{tarask01}. In this respect, it is
worth noting that the transverse acoustic branch at $Q$ larger
than $\approx$8 nm$^{-1}$ is the glassy counterpart of a
transverse optic phonon branch of $\alpha$-quartz (almost flat at
$\approx$4 THz, i.~e. $\approx$16 meV). This branch, in the
extended Brillouin zone scheme which is  more appropriate for
disordered materials, is the prosecution of the transverse
acoustic branch \cite{dorner}. According to Boysen at al.
\cite{boysen}, this branch is (close to the M-point) strongly
temperature-dependent, and its softening is responsible for the
$\alpha$-$\beta$ transition in quartz. More importantly, the
atomic displacements induced by the lattice modes of this branch
in quartz, as determined in \cite{dorner}, are very similar to the
frustrated localized rotation of SiO$_4$ tetrahedra, that in
vitreous silica -according to Buchenau et al. \cite{buc86}- are
the modes contributing to the BP.

The assignment of the BP to the flattening of the transverse
acoustic branch is strengthened by the behaviour of the high-energy
transverse dynamics upon densification. Spectra similar to those
of Fig.~2, but for the sample at the highest studied density
($\rho$=4.0 g/cm$^3$), are reported in Fig.~4, while the
corresponding dispersion curves are reported in Fig.~3. One can
observe that (i) the two current spectra are now "pure", and no
evidence of wrong-polarized modes is present; (ii) the positive
dispersion of the L-branch is absent; and, more importantly, (iii)
the T-branch no longer shows a flattening. This latter
observation, together with the experimental and numerical finding
that the BP intensity strongly decreases on increasing density
\cite{inamura,inabuche,jund} (as evident in Fig. 1), gives a
decisive support to the finding that the BP is produced by the
flattening of the quasi-TA branch.

In conclusion, by  comparing the spectra of the longitudinal and
transverse current, and by studying their density dependence, we
have shown that the BP in v-SiO$_2$ is to be ascribed to the
quasi-transverse acoustic modes, whose dispersion relation becomes
$Q$-independent at high $Q$, and consequently, gives rise to an
excess of modes with respect to the Debye behaviour. The whole
picture presented here reconciles different previous studies on
the origin of the BP: i) the high $Q$ part of the transverse
acoustic branch in vitreous silica -which gives rise to the BP- is
the counterpart of a low lying transverse optic branch in
$\alpha$-quartz \cite{dorner}; ii) according to Boysen at al.
\cite{boysen} the softening of this branch at the M-point give
rise to the $\alpha$-to-$\beta$ transition; iii) in agreement with
i) and ii), according to Taraskin et al.
\cite{tarask01,elliotpm02}, the BP arises from the softening of
the lowest-energy van Hove singularity of the corresponding
crystals; iv) according to Dorner and coworkers the eigenvectors
of this branch in quartz correspond to rotation of SiO$_4$
tetrahedra; v) finally, in agreement with i) and iv), according to
Buchenau et al. \cite{buc86} the modes contributing to the BP are
localized rotation of SiO$_4$ tetrahedra.
\par
This work was supported by INFM Iniziativa di Calcolo Parallelo,
and by MURST Progetto di Ricerca di Interesse Nazionale. One of us
(GR) greatly acknowledge illuminating discussions with U.~Buchenau
and B.~Dorner.


\clearpage
\newpage
\begin{center} 
{FIGURE CAPTIONS}
\end{center}

FIG. 1: Normalized density of states $g(E)/E^2$  of
vitreous silica at three different densities $\rho$: (a) 2.2
g/cm$^3$ (full line and circles); (b)  2.7 g/cm$^3$ (dashed line
and squares); (c) 4.0 g/cm$^3$ (triangles). Lines are
experimental data from Ref.~\cite{inabuche}, symbols are the
results of the present simulation.

\vspace{.9cm}

FIG. 2: Selected examples of longitudinal (full
lines) and transverse (dashed lines) current spectra at the
indicated $Q$-values in nm$^{-1}$, for the sample at $\rho$=2.2
g/cm$^3$.

\vspace{.9cm}

FIG. 3: Dispersion relation of the main peaks
appearing in the current spectra. Crossed circles are the maxima 
of the longitudinal current at $\rho$=2.2 g/cm$^3$ 
(the line is a guide for the eyes).
The open symbols refer to the maxima of the transverse current at 
$\rho$=4.0 g/cm$^3$ (down-triangles), 
$\rho$=2.7 g/cm$^3$ (squares), 
$\rho$=2.2 g/cm$^3$ (up-triangles).
Full symbols are INS (squares)
and IXS (circles and diamonds) experimental data for $\rho$=2.2
g/cm$^3$ taken from Ref.s \ \cite{ben96,fontana,righetti}.

\vspace{.9cm}

FIG. 4: Selected examples of longitudinal (full
lines) and transverse (dashed lines) current spectra at the
indicated $Q$-values (in nm$^{-1}$) for the highest density sample
($\rho$=4.0 g/cm$^3$).

\clearpage
\newpage
\begin{center}
{\Large Figure 1. Pilla et al.}
\end{center}

\begin{figure}
\includegraphics[width=.8\textwidth]{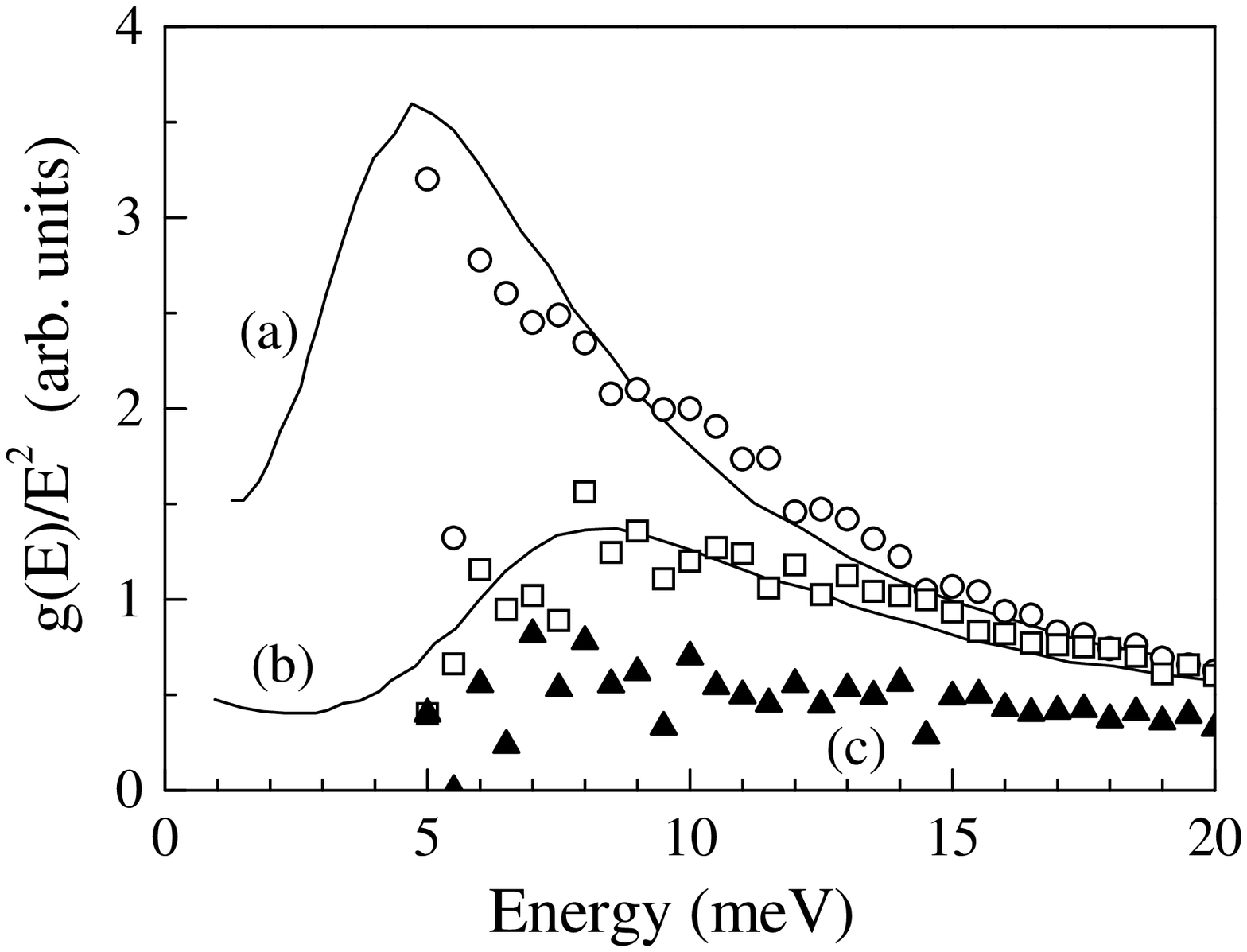}
\label{BP}
\end{figure}

\clearpage
\newpage
\begin{center}
{\Large Figure 2. Pilla et al.}
\end{center}

\begin{figure}
\includegraphics[width=.8\textwidth]{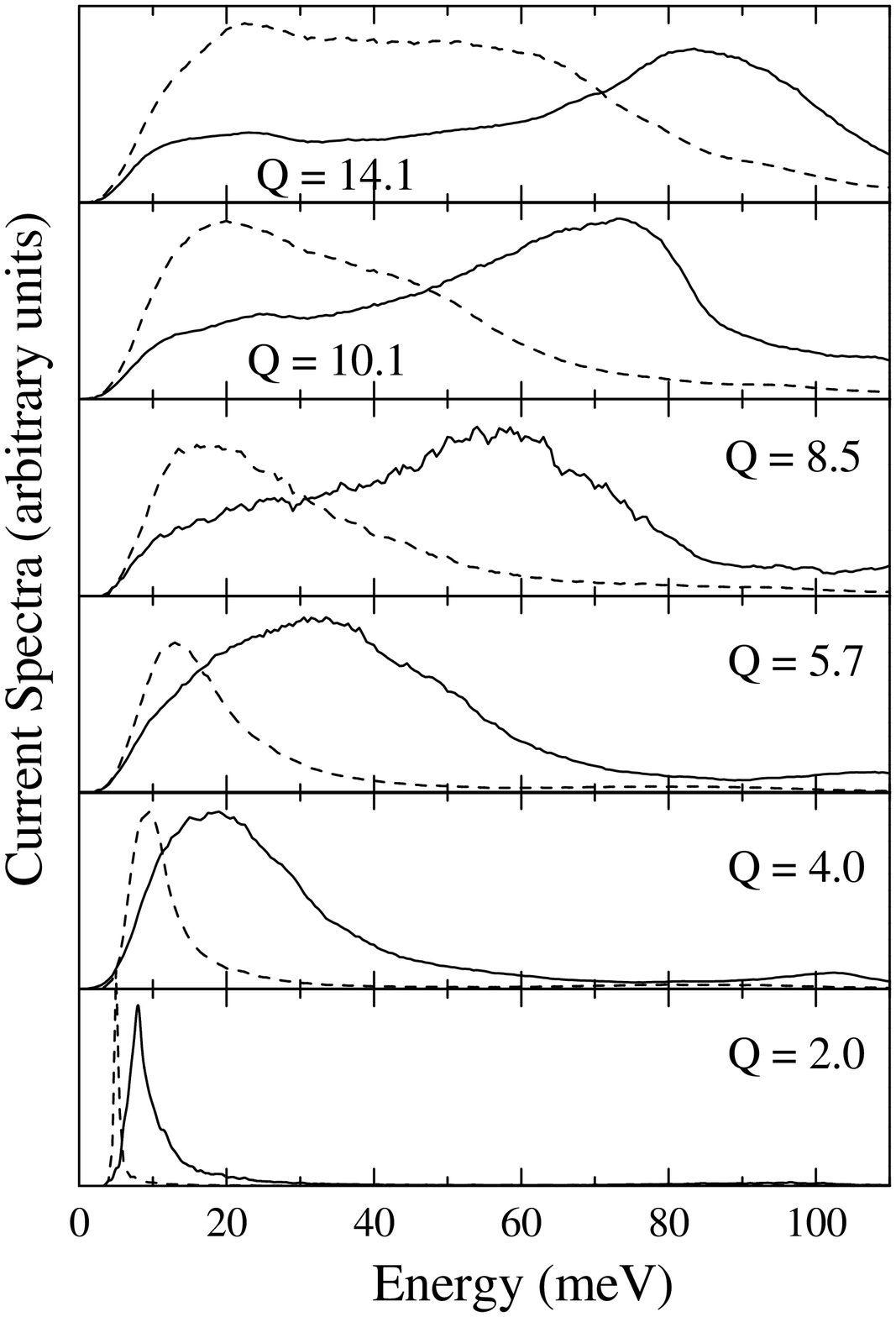}
\label{spectra}
\end{figure}

\clearpage
\newpage
\begin{center}
{\Large Figure 3. Pilla et al.}
\end{center}

\begin{figure}
\includegraphics[width=.8\textwidth]{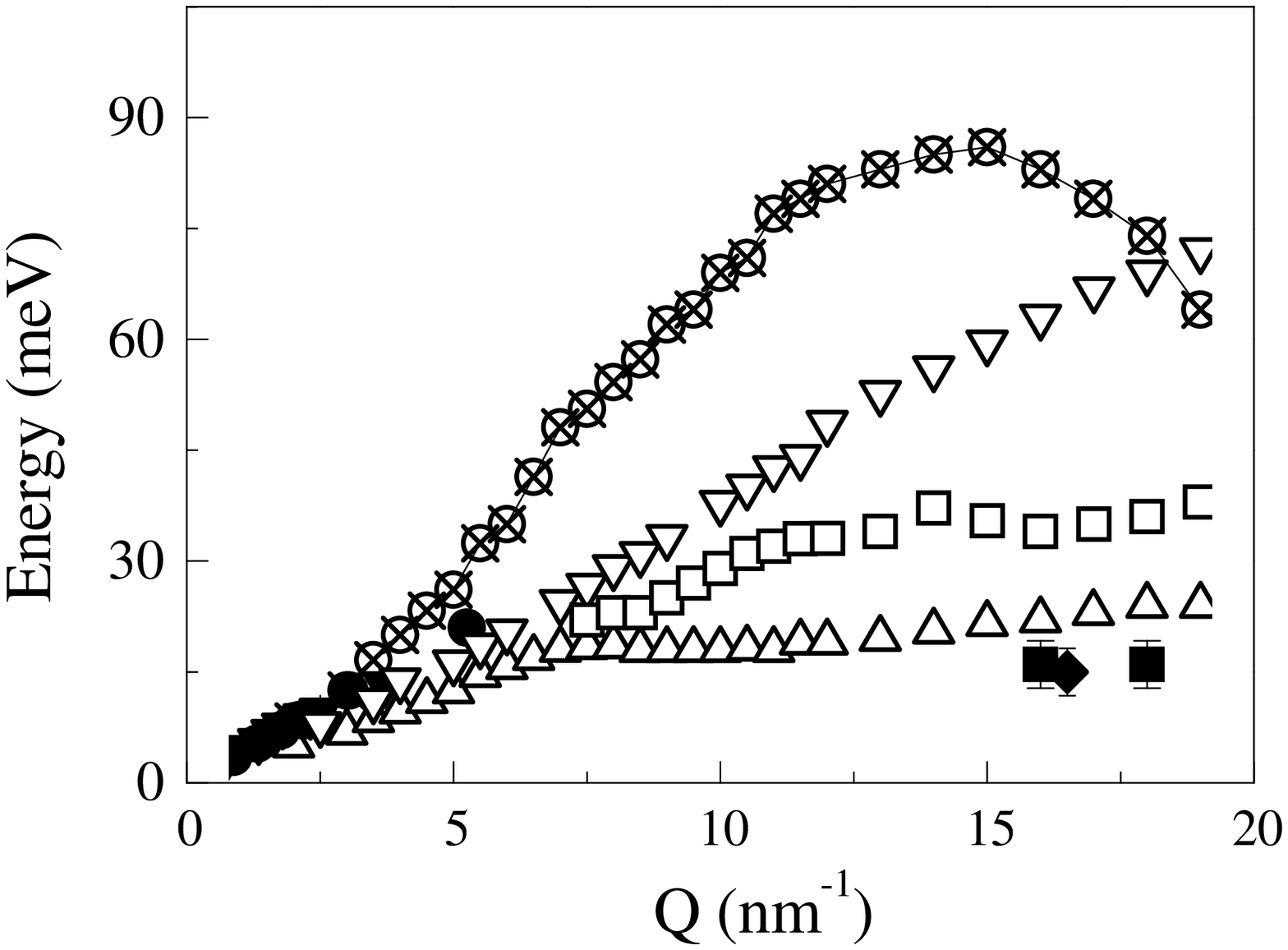}
\label{disp}
\end{figure}

\clearpage
\newpage
\begin{center}
{\Large Figure 4. Pilla et al.}
\end{center}

\begin{figure}
\includegraphics[width=.8\textwidth]{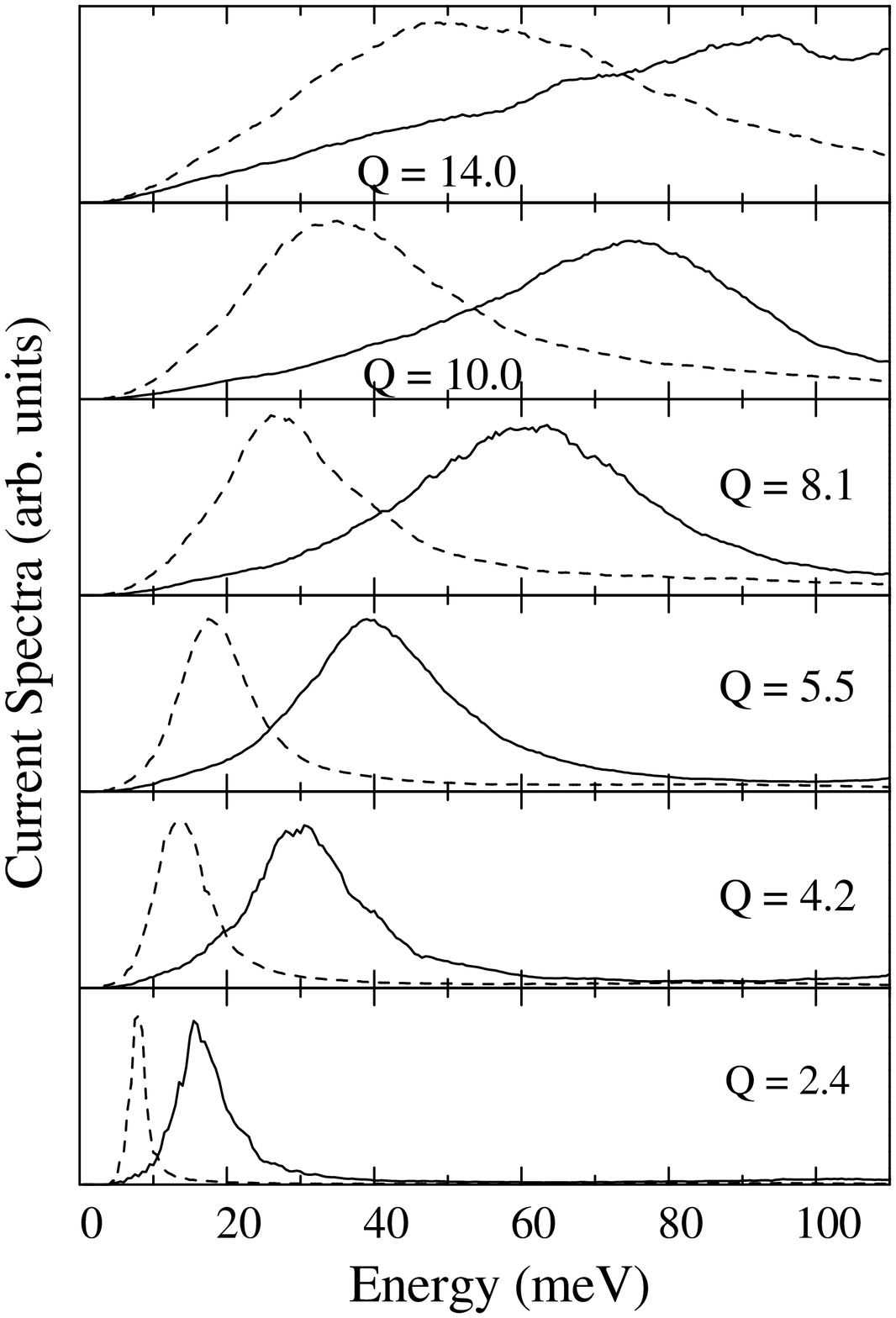}
\label{spectraHD}
\end{figure}


\begin{thebibliography}{99}
\bibitem{nota2} The dispersion relations were built by associating to each
$Q$ the energy of the peaks determined by plots like the ones in Figs. 2
and 4. For the transverse excitations, to take the peak in the transverse
currents or the low-energy peak of the longitudinal current, gives the
same results.
\end{thebibliography}
\end{document}